\documentclass[11pt]{article}
\usepackage{moriond,epsfig}

\def\Journal#1#2#3#4{{#1} {\bf #2}, #3 (#4)}

\def\NPB{{\em Nucl. Phys.} B}
\def\PLB{{\em Phys. Lett.}  B}

\def\PRD{{\em Phys. Rev.} D}

\def\EJC{{\em Euro. Phys. Journal} C}
\def\EJDC{{\em Euro. Phys. J. Direct} C}
\def\CPC{{\em Comp. Phys. Comm.}}

\def\kp{{\ifmmode{k_{\perp}}\else{$k_{\perp}$}\fi}}
\def\etmean{\ifmmode{\bar{E}^{\mathrm{jet}}_{\mathrm{T}}}\else
  {$\bar{E}^{\mathrm{jet}}_{\mathrm{T}}$}\fi}
\def\lxg{\ifmmode{\mathrm{log_{10}}(x_{\gamma})}\else
  {${\mathrm{log_{10}}(x_{\gamma})}$}\fi}
\def\xg{\ifmmode{x_{\gamma}}\else{${x_{\gamma}}$}\fi}
\def\xgp{\ifmmode{x_{\gamma}^+}\else{${x_{\gamma}^+}$}\fi}
\def\xgm{\ifmmode{x_{\gamma}^-}\else{${x_{\gamma}^-}$}\fi}
\def\xgpm{\ifmmode{x_{\gamma}^{\pm}}\else 
  {${x_{\gamma}^{\pm}}$}\fi}
\def\sqee{\ifmmode{\sqrt{s}_\mathrm{ee}}\else
  {$\sqrt{s}_\mathrm{ee}$}\fi}
\def\ipb{\ifmmode {\mathrm{pb}^{-1}}\else 
  {$\mathrm{pb}^{-1}$}\fi}
\def\grvnlo{GRV-G\,HO}

\def\be{\begin{equation}}
\def\ee{\end{equation}}
\def\bea{\begin{eqnarray}}
\def\eea{\end{eqnarray}}

\begin{document}
\vspace*{4cm}
\title{HOW WELL DOES QCD WORK FOR PHOTON-PHOTON COLLISIONS?}

\author{ THORSTEN WENGLER }

\address{CERN, EP-division, 1211 Geneva 23, Switzerland}

\maketitle\abstracts{
The performance of QCD in describing hadronic photon-photon collisions 
is investigated in the light of recent measurements from LEP on di-jet
production, light hadron transverse momentum spectra, and heavy quark
production.
}

\section{Jet Production}

\begin{figure}[t]
\psfig{figure=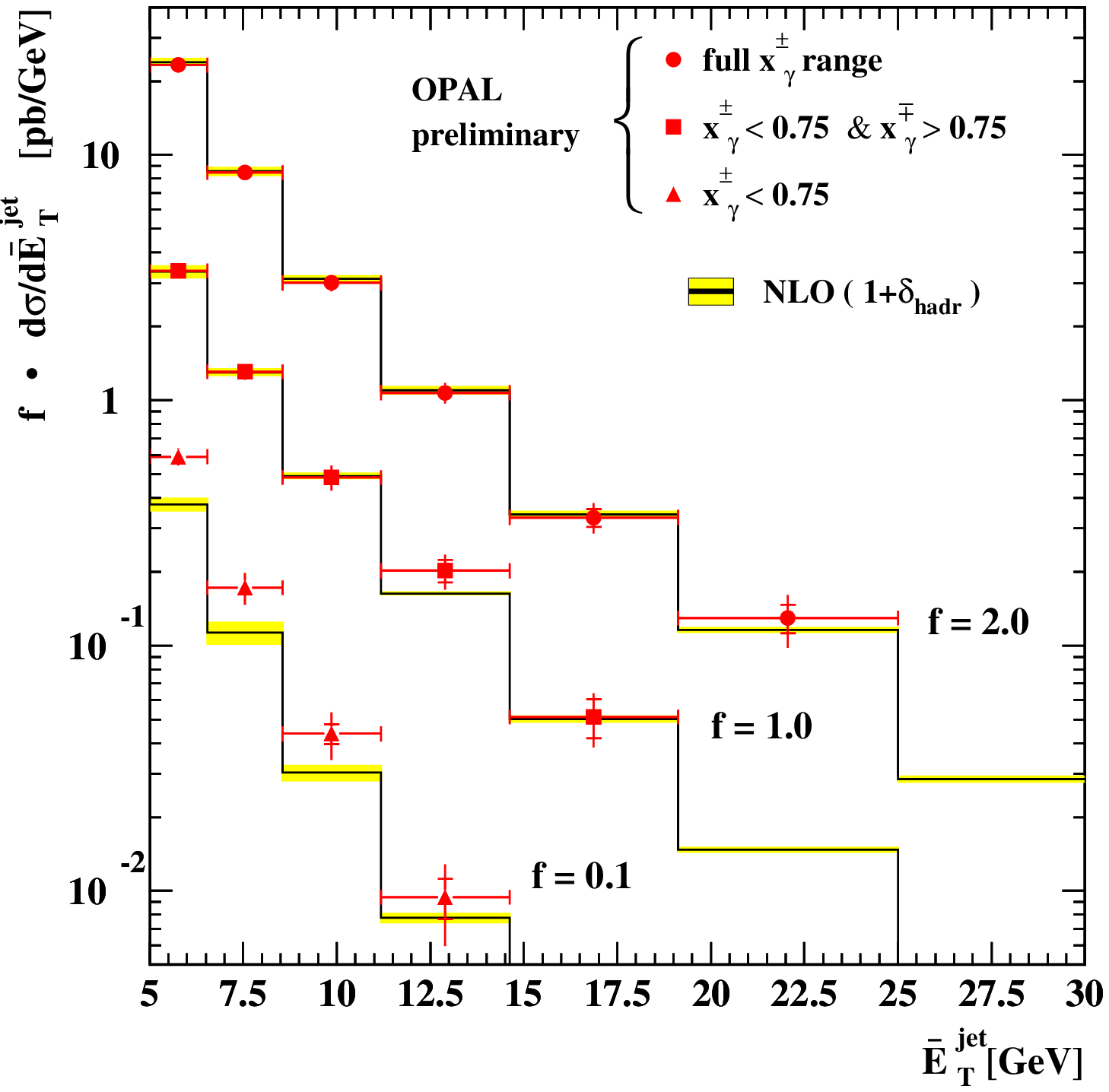,width=0.49\textwidth}
\psfig{figure=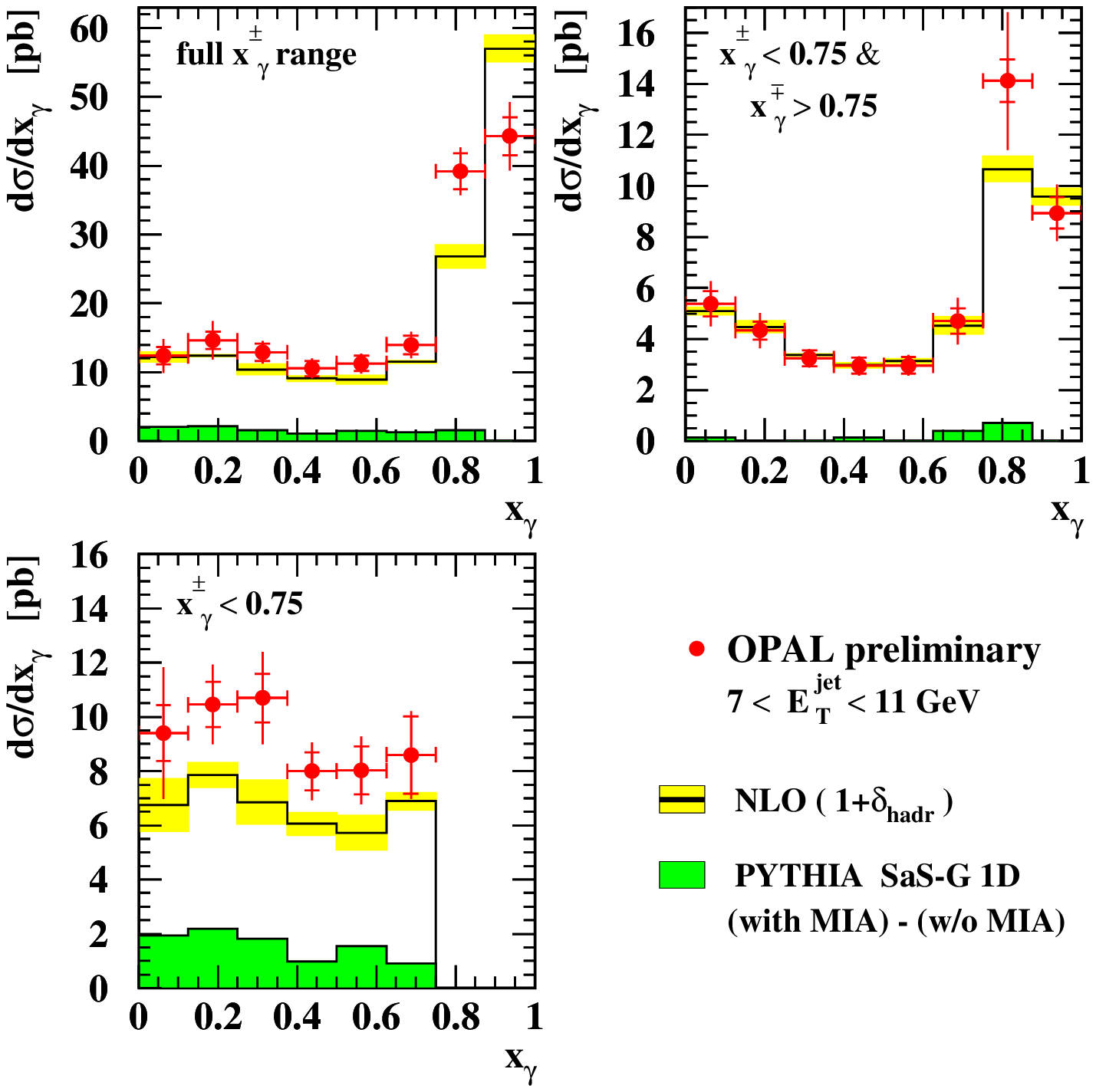,width=0.49\textwidth}
\caption{Differential di-jet cross sections as measured by OPAL.}
\label{fig1}
\end{figure}

Differential di-jet cross sections have been measured by
DELPHI~{\cite{bib-delphi-dijet}} and OPAL~{\cite{bib-opal-dijet}}.
DELPHI has analysed 220~{\ipb} of data taken at {\sqee}=192-202~GeV
using a cone algorithm with R=1 to define the jets. The first
preliminary results of the  di-jet cross sections as a function of the
jet transverse energy and the jet pseudorapidity are found to be
consistent with a previous measurement by
OPAL~{\cite{bib-opal-old-dijet}}. OPAL has used 593~{\ipb} from
{\sqee}=189~GeV to 209~GeV to study the performance of NLO
perturbative QCD and the hadronic structure of the photon in di-jet
production. The jets are reconstructed using an inclusive {\kp}
clustering algorithm~{\cite{bib-ktclus}}.
The two jets are used to estimate
the fraction of the photon momentum participating in the hard
interaction as
$ \xgpm \equiv {{\sum_{\rm jets=1,2}
(E^\mathrm{jet}{\pm}p_z^\mathrm{jet})}}/
 {{\sum_{\rm hfs}(E{\pm}p_z)}} $
where $p_z$ is the momentum component along the $z$ axis of the
detector and $E$ is the energy of the jets or objects of the
hadronic final state (hfs). In single (double) resolved processes,
where one (both) of the photons interacts via a fluctuation into a
hadronic state, the hard interaction is accompanied by one or two
remnant jets, and one or both {\xg} are smaller than one.
Differential cross sections as a function of {\xg} or in regions of
{\xg} are therefore a sensitive probe of the structure of the photon.
The left plot in Figure~{\ref{fig1}} shows the differential di-jet
cross section as a function of the mean transverse energy {\etmean} of
the di-jet system for the full {\xg}-range, for either {\xgp} or
{\xgm}$<$1 (dominated by single resolved processes), or {\xgpm}$<$1
(dominated by double resolved processes). The prediction of
perturbative QCD in NLO~\cite{bib-ggnlo} using the 
{\grvnlo}~\cite{bib-grv} parton densities is compared to the data
after hadronisation corrections have been applied to the calculation.
The calculation is in good agreement with
the data, except for being to low at small {\etmean} for {\xgpm}$<$1.
In this region the contribution of the so called underlying event as
described by the concept of multiple parton interactions (MIA) is
expected to be largest. This contribution is not included in the NLO
calculation. The three plots on the right hand side of
Figure~{\ref{fig1}} show the differential cross section as a function
of {\xg} for the three regions in {\xgp}-{\xgm}-space described
above. The shaded histogram on the bottom of each of the three plots
indicates the contribution of MIA to the cross section as obtained
from the PYTHIA~{\cite{bib-pythia}} MC generator. It is evident
especially for {\xgpm}$<$1 that
the MIA contribution is of about the same size as the discrepancy 
between the measurement and the NLO prediction. Furthermore it is
interesting to observe that there is next to no MIA contribution
to the cross section if either {\xgp} or {\xgm} is required to be less
than one, while the sensitivity to the photon structure at small {\xg}
is retained. As one would expect also the agreement of the NLO calculation
with the measurement is best in this case. With these measurements one
is therefore able to disentangle the hard subprocess from soft
contributions and make the firm statement that NLO perturbative QCD is
adequate to describe di-jet production in photon-photon collisions. At
the same time a different sub-set of observables can be used to study
in more detail the nature of the soft processes leading to the
underlying event.

\section{Transverse Momentum Spectra of Light Hadrons}

\begin{figure}[t]
\psfig{figure=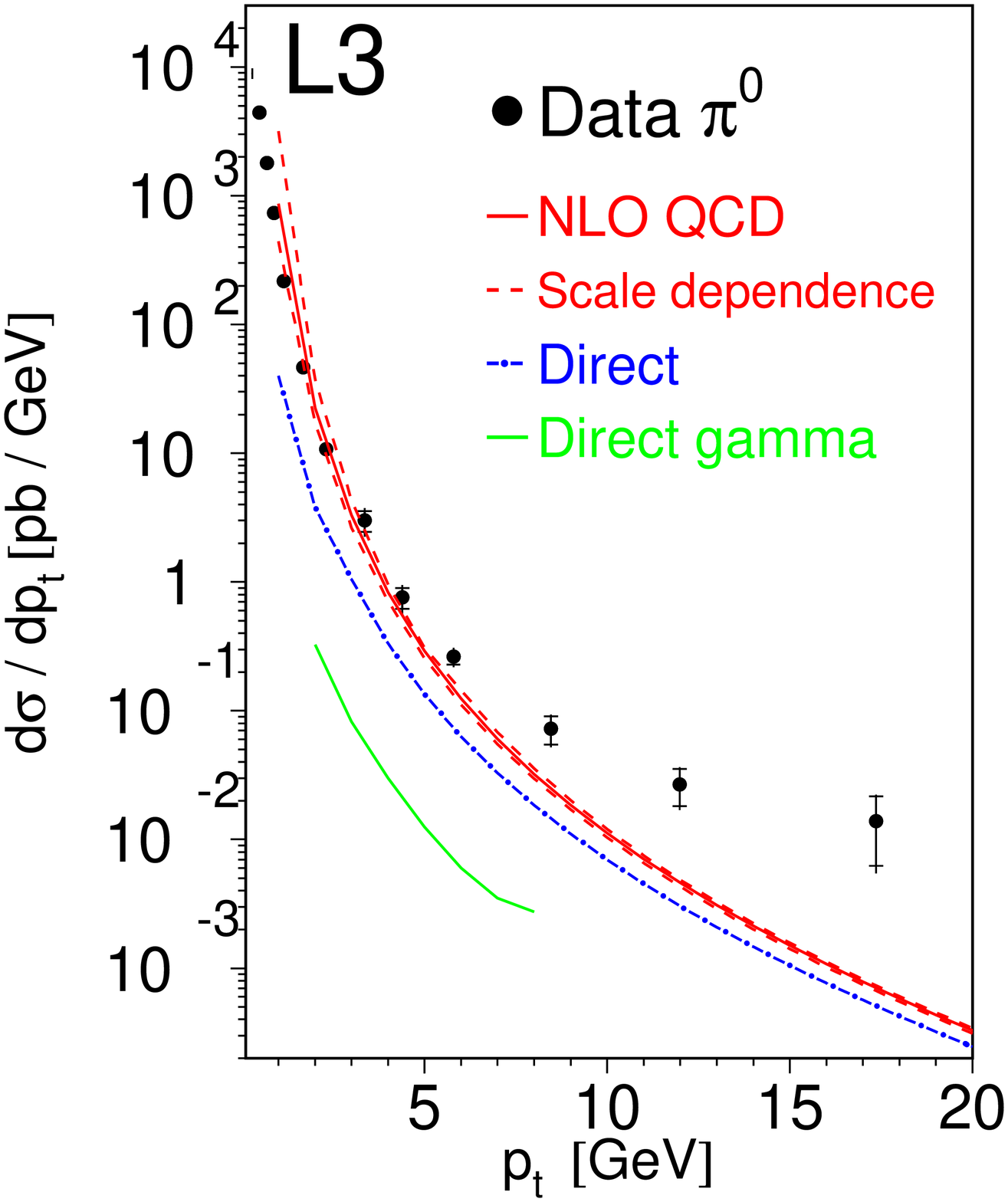,width=0.42\textwidth}
\psfig{figure=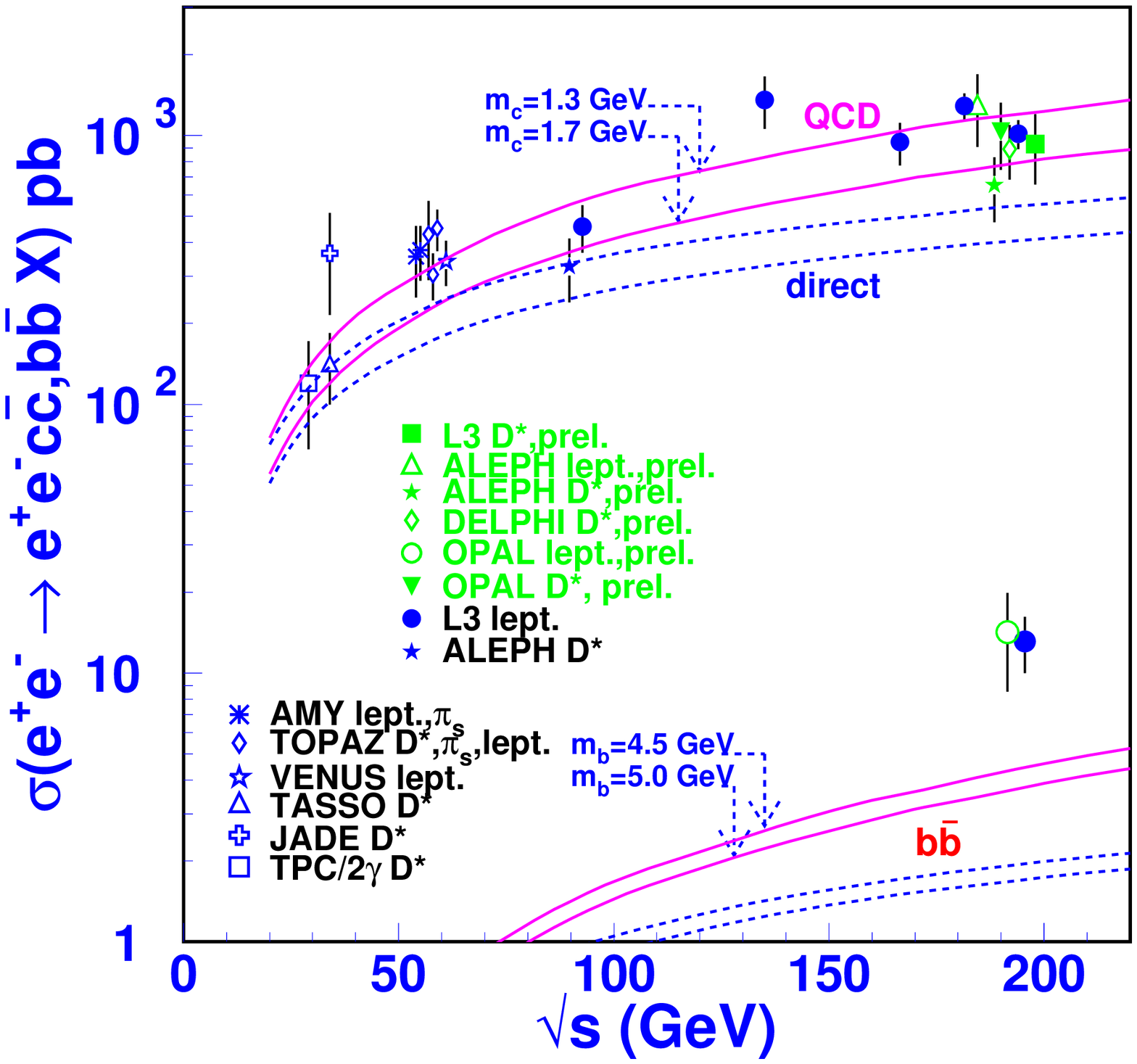,width=0.56\textwidth}
\caption{The differential cross section as a function of the
  transverse momentum of the hadron for neutral pions as measured by
  L3 (left) and the total cross section of charm production (ALEPH,
  DEPHI, L3, OPAL) and beauty production (L3, OPAL) in photon-photon
  collisions (right).}
\label{fig2}
\end{figure}

The L3 collaboration has recently presented new results on the
inclusive production of K$^0_S$ and neutral and charged pions in
414~{\ipb} of data taken at {\sqee} from 189~GeV to
202~GeV~{\cite{bib-l3-inclhad}}. Compared to the jet measurements
described above, inclusive hadron production offers a
complementary signature to study the dynamics in hadronic photon-photon
collisions. Predictions are available in NLO perturbative QCD which
make use of fragmentation functions to translate the partonic cross
sections to the observables measured. The uncertainty associated with
the hadronisation correction commonly applied to the calculation of
partonic NLO jet cross sections can hence be avoided, but is of
course replaced by any uncertainty attached to the determination of the
fragmentation functions. The new measurements can be shown to be
consistent with results obtained previously by the OPAL
collaboration~{\cite{bib-opal-had}} where both measurements overlap in
phase space. The new L3 data significantly extends the older
measurements towards high transverse momenta for the production of
charged and neutral pions. As can be seen in Figure~{\ref{fig2}} it is in
this region that a significant discrepancy between the measurement and
the QCD prediction occurs. The left plot in Figure~{\ref{fig2}} shows
the differential cross section for inclusive {$\pi^o$}-production as a
function of the transverse momentum $p_T$ of the {$\pi^o$}. For $p_T>$
8~GeV the data exhibits a significantly harder spectrum than predicted
by NLO QCD. This behaviour is confirmed by the
independent measurement of charged pions. 
This discrepancy is particularly remarkable as it
occurs at high transverse momenta of the produced hadrons, which in
turn suggests an underlying high momentum partonic process. Under
these conditions one expects the perturbative calculation to be
reliable due to the relatively small value of the strong coupling
constant. The discrepancy also appears to be in contradiction with the
good agreement of the NLO QCD calculation obtained for the jet measurements
described above, which should be sensitive to a similar set of
processes and mainly differs in that fragmentation functions are not
needed here. A possible direction for a further investigation of this
anomaly may be derived from the OPAL
measurement~{\cite{bib-opal-had}}. 
Here the data is presented in four regions of
the invariant hadronic mass $W$. For $W<$ 50~GeV a similar tendency of
a harder $p_T$ spectrum in the data then in the theory can be
observed. For $W>$ 50~GeV, however, the agreement is good. It would
therefore be interesting to repeat these measurements for the full
LEP2 data set in several regions of $W$.

\section{Heavy Flavour Production}

Yet another test of the performance of perturbative QCD in
photon-photon collisions is the production of heavy quarks. Due to the
large physical scale set by the charm or beauty mass one can expect
the theoretical prediction form perturbative QCD to be reliable. A
significant part of the cross section at LEP2 energies is predicted to
be from resolved processes, in particular the photon-gluon fusion
{$\gamma g \rightarrow c\bar{c} (b\bar{b})$}. The production rate of
charm and beauty quarks therefore depends not only on their mass but
also on the gluon density in the photon. The latest results from the
LEP collaborations on charm production use identified charged D$^*$
meson for charm tagging. L3~{\cite{bib-l3-cc}} has analysed 683~{\ipb}
at {\sqee}=183-209~GeV, DELPHI~{\cite{bib-delphi-cc}} 458~{\ipb} at
{\sqee}=183-209~GeV, OPAL~{\cite{bib-opal-cc}} 428~{\ipb} at
{\sqee}=183-202~GeV, and ALEPH~{\cite{bib-aleph-cc}} 236~{\ipb} at
{\sqee}=183-189~GeV. The 
smallest uncertainties are to be expected if one compares the
measurements and the theoretical predictions for the restricted phase
space corresponding to the experimental acceptances, as any
extrapolation to the total charm production cross section introduces
additional assumptions. The differential cross section of charged
D$^*$ production as a function of the D$^*$ transverse momentum is in
good agreement among the four experiments, with the possible exception
of a slightly harder spectrum for ALEPH. NLO perturbative QCD using
massive quarks~{\cite{bib-cc-nlo}} and a charm mass of m$_c$=1.5~GeV
describes the data well. This translates to similar agreement
of the total charm production cross section with the QCD
prediction~{\cite{bib-cc-nlo}} as can be seen in the right plot of
Figure~{\ref{fig2}}. In the same figure the results are shown for
the total production cross section of b$\bar{\mathrm{b}}$ as obtained
by the 
L3 and the OPAL collaborations. Both measurements are based on the
higher transverse momentum of the lepton in semileptonic b-decays
as compared to those of lighter quarks. L3 has analysed 410~{\ipb}
at {\sqee}=189-202~GeV using both the electron and the muon
signatures~{\cite{bib-l3-bb}}, OPAL has analysed 371~{\ipb} at
{\sqee}=189-202~GeV using the muon channel~{\cite{bib-opal-bb}}. The
plot demonstrates that there is good agreement between the
experimental results. The prediction of perturbative QCD in NLO
however significantly underestimates the measurements. 
Similar deficiencies have been observed in b-production in
p$\bar{\mathrm{p}}$- and ep-collisions. A recent analysis of the
b-fragmentation functions used in these calculations suggests
that much of this discrepancy can be recovered at least for
p$\bar{\mathrm{p}}$-collisions~{\cite{bib-cacciari}}. This analysis
considers also the higher moments of the differential distributions of
the average b-quark energy fraction carried by the B-meson. Using
recent LEP and SLD data in this way to fix the b-fragmentation the
agreement of the perturbative calculation with the
p$\bar{\mathrm{p}}$-data becomes acceptable. One might hope that a
similar technique applied to ep- and photon-photon-collisions will
also improve the agreement of the theory with the measurements.

\end{document}